# Sentiment Analysis of Political Tweets for Israel using Machine Learning

Amisha Gangwar[1] and Tanvi Mehta [2]

[1] Machine Learning Researcher at LearnByResearch, Bareilly
[2] Machine Learning Researcher at LearnByResearch, Pune
[1]amishagangwar21@gmail.com
[2]tanvihm11@gmail.com

**Abstract.** Sentiment Analysis is a vital research topic in the field of Computer Science. With the accelerated development of Information Technology and social networks, a massive amount of data related to comment texts has been generated on web applications or social media platforms like Twitter. Due to this, people have actively started proliferating general information and the information related to political opinions, which becomes an important reason for analyzing public reactions. Most researchers have used social media specifics or contents to analyze and predict public opinion concerning political events. This research proposes an analytical study using Israeli political Twitter data to interpret public opinion towards the Palestinian-Israeli conflict. The attitudes of ethnic groups and opinion leaders in the form of tweets are analyzed using Machine Learning algorithms like Support Vector Classifier (SVC), Decision Tree (DT), and Naïve Bayes (NB). Finally, a comparative analysis is done based on experimental results from different models.

## 1 Introduction

With the rapid growth of Online Social Networks (OSNs), communication platforms have become popular. This has helped large numbers of the population share, search, and interchange data and information without any concern about geographical distance [1]. The volume of data created through social media platforms, especially Twitter, is massive. Twitter is an online news and social networking site where people communicate in short tweets [2]. It has become the most popular social media platform, with millions of users posting tweets each day.

The data relating to public opinion has increased tremendously [3]. Identification of these opinions regarding political events or issues is essential to form international alliances, policies, and positions. The actions of government officials depend on these opinions, so they should keep an eye on this data to make future decisions [4]. Opinion

---

polls have been the standard mechanism for collecting public opinion, often giving rise to several problems. It is not easy to interpret fine-grained analysis or learn the intentions, subjectivity, and motivations behind public opinion with polls [5]. All the aforestated drawbacks make opinion polls not very well-grounded, and there is a requirement for advanced mechanisms to understand such public views.

The massive number of users, the diversified topics, and enormous volumes of posted tweets or content have resulted in social media becoming a rich source to predict the population's attitudes [6]. Mining or using advanced techniques for political opinions may provide a faster, more accurate, and less expensive alternative to traditional polls [7]. Several research works have explored social media mining to analyze and predict public political opinions. However, these research works were mostly event-specific and used only relevant techniques to investigate the issue [8]. Furthermore, most of the studies depend on sentiment analysis to predict the user's feelings rather than the political opinion.

In politics, analyzing a sentiment depends on the side one stands by regardless of the users' views [9]. Hence, a sentiment analyzer may classify a tweet, phrase, comment, or idiom as "negative" or "positive." Furthermore, few researchers have analyzed current situations statistically instead of making predictions about public opinions [10]. This paper proposes an analytical and comparative study of the Twitter text dataset to examine the public political opinion in several countries towards the Israelis in the Palestinian-Israeli conflict. The data worked upon is dated May 2021. The proposed method uses the machine learning model based on Support Vector Classifier (SVC), Decision Tree (DT), and Naïve Bayes (NB) algorithms, and the results show a comparative analysis of these algorithms.

In section 2, a summary of the previous research works is presented. In section 3, the techniques used in the proposed research are briefly described. Section 4 describes the implementation of the work. The analytical outcomes and evaluation are tabulated in section 5. The conclusions and future scope are discussed in section 6.

## 2 Literature Review

Bhawani Selvaretnam et al., in 2017, [11] designed an experiment that extracts the sentiments based on the subjects in tweets using NLP (Natural Language Processing). This was done in three steps: classification, semantic association, and polarity sort by identifying the grammatical relationship between subject and sentiment lexicon. SVM technique resulted in better accuracy.

Rajon Bardhan et al. in 2018, [12] presented sentiment analysis on user comments based on Natural Language Processing (NLP) used to generate datasets. The solution resulted in a data-driven experiment in the accuracy of finding the popular and high-quality video. The efficiency obtained through this approach was 75.44%.

Risky Novendri et al. in 2020 [13] presented a solution for analyzing viewers' comments on Netflix series episodes through sentiment analysis using the Naïve Bayes algorithm. The results were 81%, 74.83%, and 75.22% for accuracy, precision, and recall, respectively.

Hoda Korashy et al., in 2014, [14] has done a comprehensive survey of sentiment analysis. Most of the proposed research and algorithms are being investigated. They survey various methods like transfer learning emotion detection, including sentiment analysis. Nhan Cach Dang et al. in 2021, [15] reviewed the study of sentiment analysis problems solved using deep learning, like sentiment polarity. They applied TF-IDF and word embedding techniques to the datasets. The comparative analysis is done based on the experimental results. The research concluded that RNN is reliable.

Van Cuong Tran et al. in 2018 [16] explored the new approach based on a feature ensemble model containing fuzzy sentiment related to tweets by considering lexical and position of words by polarity. They experimented on actual data to study the performance of various algorithms.

Matheus Araújo et al., in 2014, [17] have compared eight sentiment analysis methods in the context of agreement and coverage. They developed a new way that combined best agreement and coverage results. They also presented iFeel, a free Web service that provides an open-access API through which compare different sentiment analysis methods.

Angelika Maag et al. in 2019 [18] showed that sentiment analysis improves granularity at the aspect level. They explored aspect extraction and sentiment classification of target-dependent tweets using deep learning techniques. A comparative study of deep learning algorithms like CNN, LSTM, and GRU, for aspect-based analysis, was done.

Apoorv Agarwal et al. in 2018 [19] performed Twitter data sentiment analysis by introducing features like POS (Specific Prior Polarity) and explored the solution for feature engineering by introducing a tree kernel that completed the unigram baseline. In a feature-based approach, they concluded that important features combine the part-of-speech and prior-polarity of words.

Xiaoyu Qiu et al., in 2018, [20] has worked on improving word representation methods. This method integrated sentiment information into the traditional TF-IDF algorithm and generated weighted word vectors. The sentiment analysis was done using RNN, CNN, LSTM, NB. The results obtained proved that the proposed method gave the highest accuracy of 92.18%.

Osama Abu-Dahrooj et al. in 2019, [21] studied data analysis models exploring country-level analysis and individual-level analysis of Palestine Tweets. The data was analyzed based on the activity and sentiments on the country and individual comments. They used a multi-level analysis technique for this approach.

Shahla Nemati et al. in 2021, [22] proposed a method for sentiment analysis using Attention-based Bidirectional CNN-RNN Deep Model (ABCDM) on five datasets. Independent bidirectional layers like LAST and GRU were used, ABCDM did the extraction of context by applying attention mechanisms on the output of layers. Polarity detection was used to calculate the effectiveness of ABCDM. The accuracy obtained by ABCDM was 96.8%.

# 3 Techniques

The proposed techniques for Twitter Sentiment Analysis are briefly explained in this section. The techniques are:

## 3.1 Support Vector Classifier (SVC)

SVC is applied for classification challenges. The data is segregated using the best decision boundary in this supervised algorithm, i.e., the hyperplane. It selects the extreme points called support vectors that help construct the hyperplane. It also has a positive hyperplane that passes through one (or more) of the nearest positive attributes and a negative hyperplane that passes through one (or more) of the nearest negative points [23]. The optimal hyperplane is where the margin, distance between positive and negative hyperplane, is maximum.

## 3.2 Decision Tree (DT)

The decision tree is not only applied for classification but also regression challenges. As the name suggests, this supervised algorithm uses the tree structure to depict the predictions that result from a series of element-based splits. It begins with the root node and ends with a decision made by leaves. It acquires basic terminologies like root nodes, leaf nodes, decision nodes, pruning, sub-tree [24]. This technique possesses a bunch of if-else statements where it checks if the condition is true; if yes, it moves forward to the next node attached to that decision.

## 3.3 Naïve Bayes (NB)

Naïve Bayes, being a supervised algorithm, is based on Bayes Theorem and is applied for classification challenges. It is mainly used in text classification, including high dimensional training datasets [25]. It helps build fast and accurate machine learning models and make quick predictions.

## 3.4 Bag of Words (BoW)

The BoW is a method of representing text data when modeling text with machine learning techniques. This model turns arbitrary text into fixed-length vectors by counting how many times each word appears. This process is known as vectorization. This simple implementation is used to solve problems related to language modeling and document classification by extracting text features. It offers much flexibility for customization on specific text data.

# 4 Methodology & Implementation

This section delineates the usual process of implementing sentiment analysis. The schematic representation of the proposed method is demonstrated in Fig. 1. A brief discussion of the flow is done below:

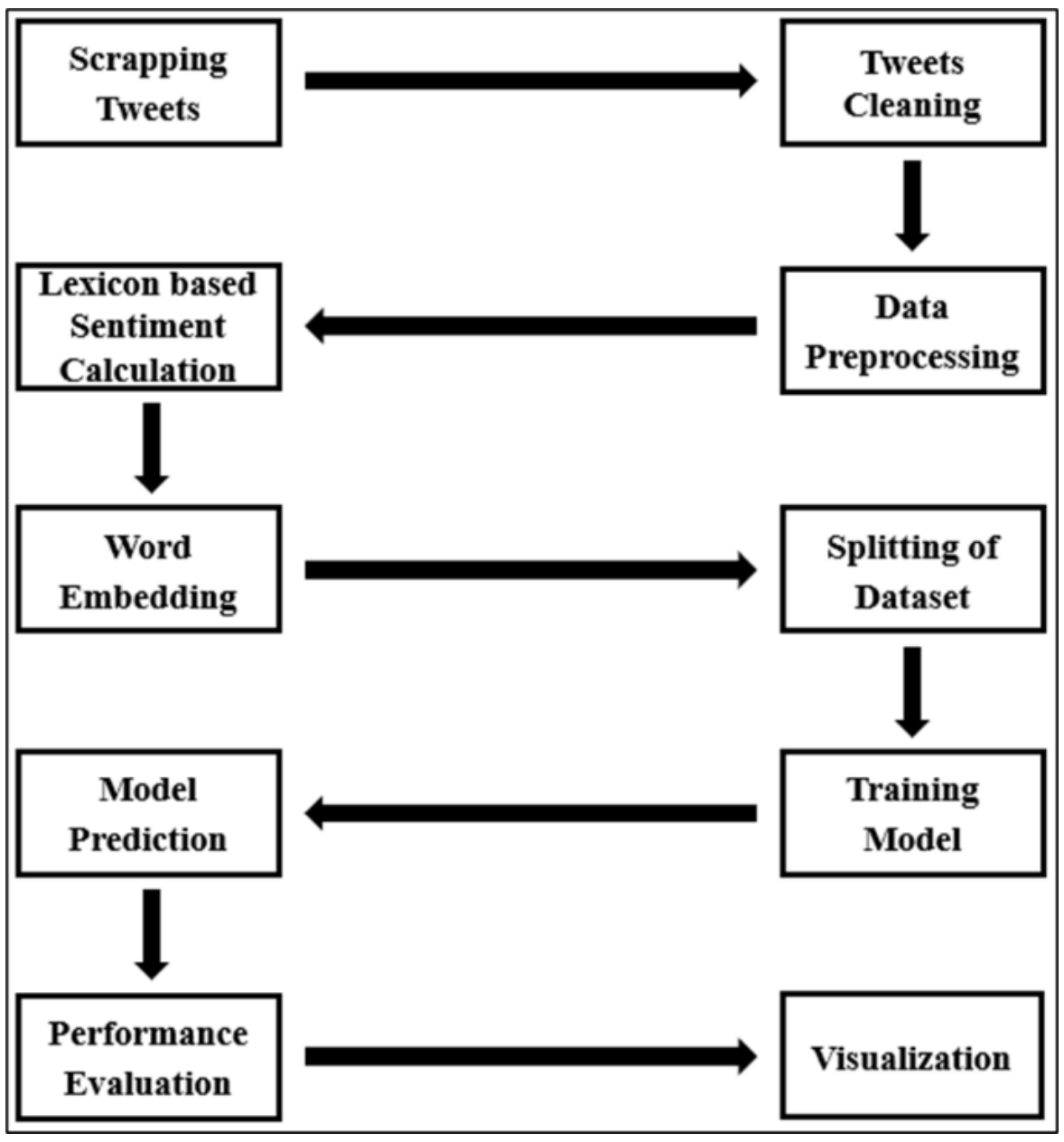

**Fig. 1.** Schematic Representation of Methodology

## 4.1 Dataset Scrapping

The dataset was scraped using Twitter Intelligence Tool (TWINT) from 6th May 2021 to 21st May 2021. 37,530 English tweets were extracted with hashtags - #IsraelUnderAttack, #IStandWithIsrael, #WeStandWithIsrael and #IsraelPalestineconflict. Fig. 2. depicts the world cloud of our collected dataset.

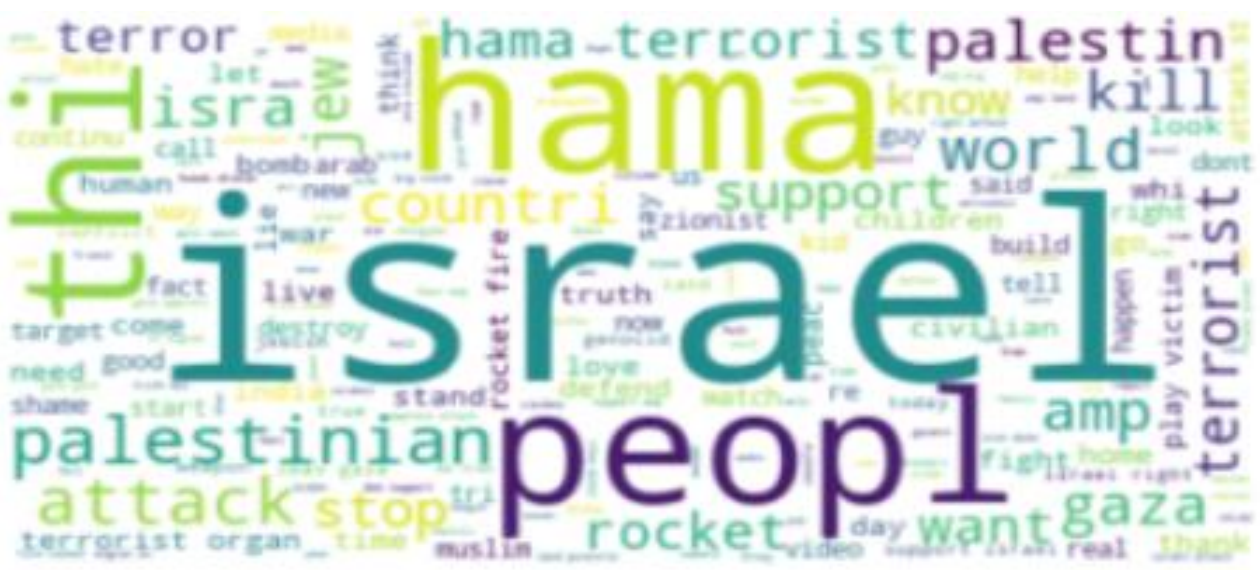

**Fig. 2.** World Cloud of Dataset

### 4.2 Dataset Cleaning

Data were cleaned by removing duplicates and rows with no tweet using drop_duplicates() and dropna(). After this process, only 32,885 tweets were left.

### 4.3 Dataset Pre-processing

In this step, data is manipulated to get and extract the essential data from it. Tweets were converted into normal text. Later, normal data was pre-processed to a suitable form by removing stopwords, extra white spaces, punctuation, and lowercasing. Now the long tweets need to be split up into words or tokens that can be processed. It was done using NLTK (Natural Language ToolKit), including a pre-trained 'punkt' English language tokenizer. Further, words are converted into base or root words by stemming. NLTK - Porter Stemmer was used to achieving this task.

### 4.4 Lexicon Based Sentiment Analysis

This step includes the calculation of sentiment from the semantic orientation of words or phrases that occur in a text. It was done using the python package TextBlob, which returns Polarity and Subjectivity. Polarity values lie between [-1,1], where -1 indicates Negative sentiment, 0 indicates neutral, and +1 is positive. Subjectivity values lie between [0,1], which tells if the text is stating a fact or personal opinion.

### 4.5 Word Embedding

Embedding is a method to extract the meaning of a word. Each word was translated to a vector using the CountVectorizer tool from the scikit-learn library.

### 4.6 Splitting Dataset

In this step, the required dataset gets bifurcated into training and testing sets. The training–the testing proportion is 80% - 20%, respectively.

### 4.7 Training Model

This step comprises implementing a classification model with the help of classifiers like SVC, DT, and NB.

### 4.8 Model Prediction

Various models like SVC, DT, and NB were used and classification reports and confusion matrix were predicted.

### 4.9 Performance Evaluation and Visualization

The accuracy metrics and confusion matrices analyzed the performance of various models. The classifier obtaining higher accuracy will be the most efficient one. The confusion matrices are plotted using a heatmap. Furthermore, a comparative analysis of these accuracies was done.

## 5 Experimental Results & Analysis

Fig. 3 shows the subjective and polarity of the cleaned tweets, which suggests our dataset contains a great range of tweets.

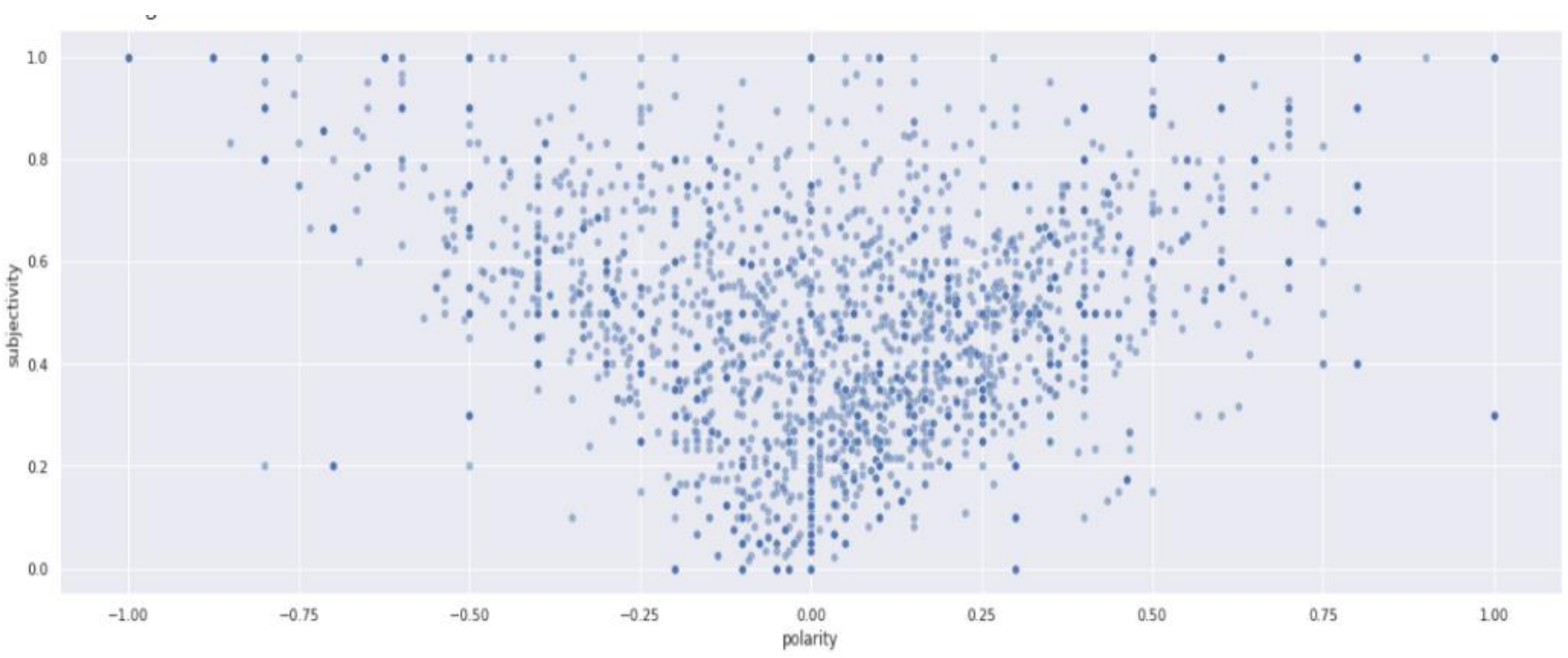

**Fig. 3.** Subjective Vs. Polarity Plot

The results achieved from all the models are precision, recall, and F1 score. The tweets are distinguished into positive, neutral, and negative. These results with accuracies are tabulated in Table 1 and Table 2. The graph of the accuracy of all models is demonstrated in Fig. 4. It shows that NB obtains the highest accuracy of 93.21%.

**Table 1.** Comparative Analysis of Results

| Models | Positive | | | Neutral | | | Negative | | |
|---|---|---|---|---|---|---|---|---|---|
| | **Precision** | **Recall** | **F1 Score** | **Precision** | **Recall** | **F1 Score** | **Precision** | **Recall** | **F1 Score** |
| SVC | 0.92 | 0.79 | 0.85 | 0.91 | 0.97 | 0.94 | 0.81 | 0.82 | 0.82 |
| DT | 1.00 | 0.96 | 0.98 | 0.91 | 0.97 | 0.94 | 0.84 | 0.67 | 0.75 |
| NB | 1.00 | 1.00 | 1.00 | 0.98 | 0.90 | 0.94 | 0.73 | 0.93 | 0.82 |

**Table 2.** Comparative Analysis of Accuracy

| **Models** | SVC | DT | NB |
|---|---|---|---|
| **Accuracy** | 0.8976 | 0. 9223 | 0.9321 |

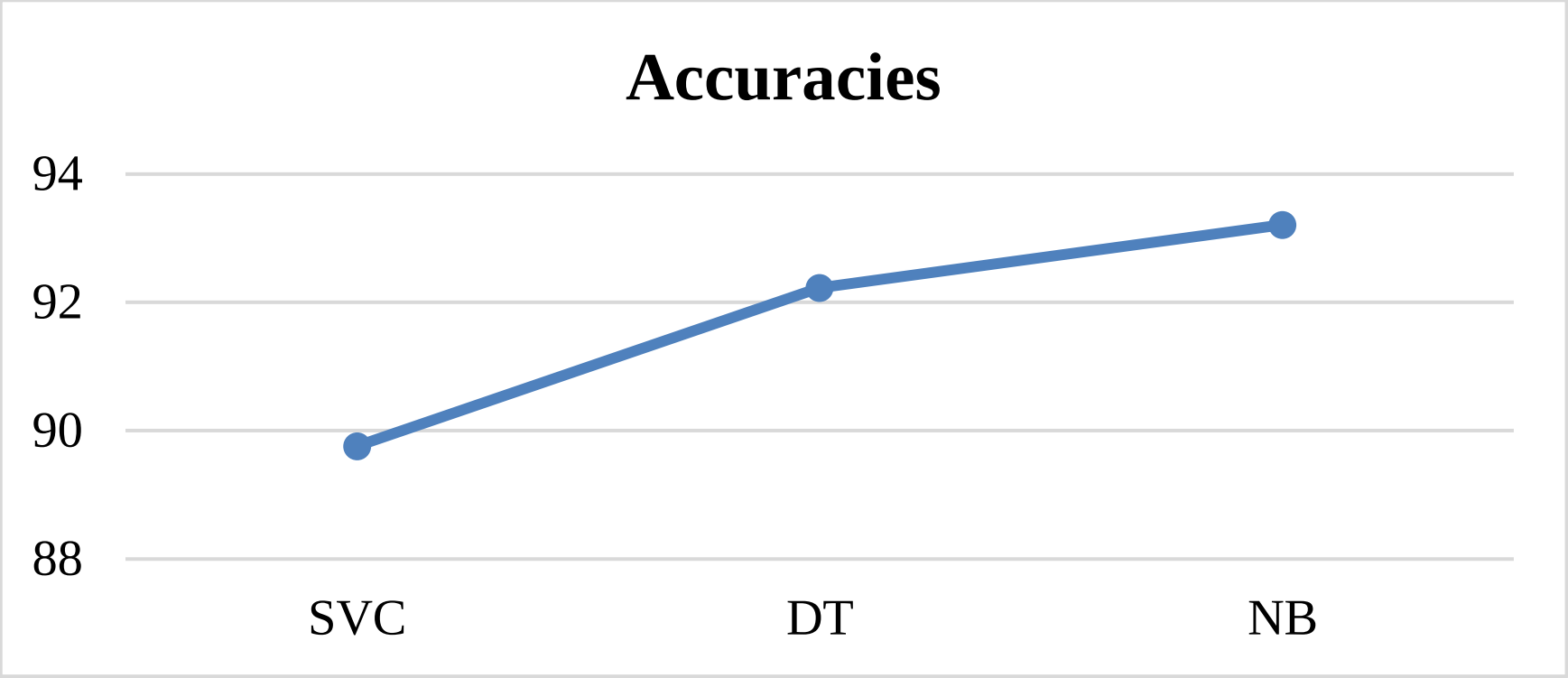

**Fig. 4.** Accuracy Plot of Models

The confusion matrix gives insights into the performance of Machine Learning models on a test set whose actual labels are known. In other words, it tells how much the model is confused about the loaded data set. The confusion matrices of SVC, DT, and NB are shown in Fig. 5, Fig. 6, and Fig. 7, respectively.

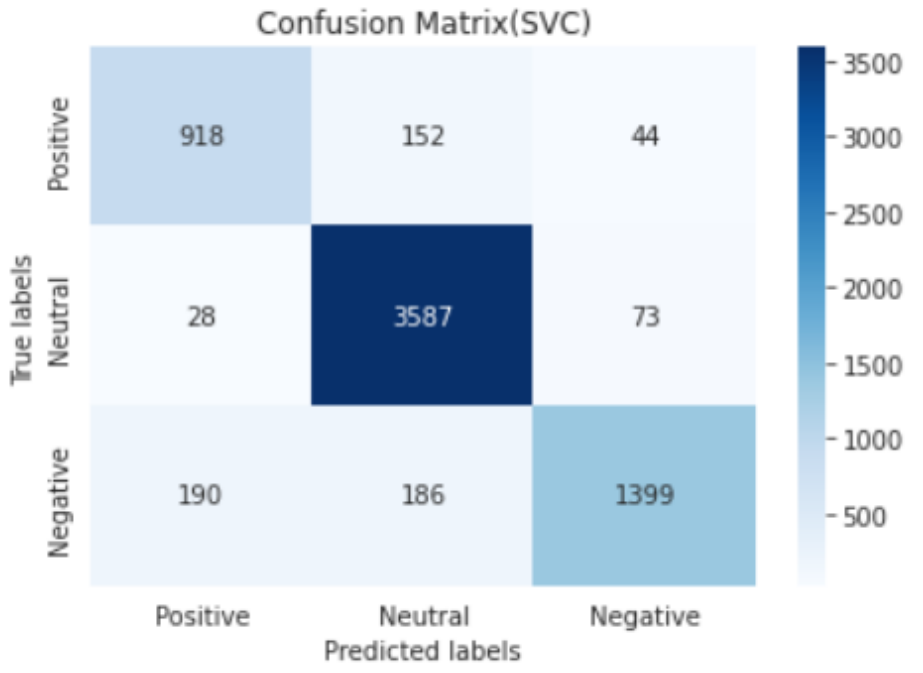

**Fig. 5.** Confusion Matrix of SVC

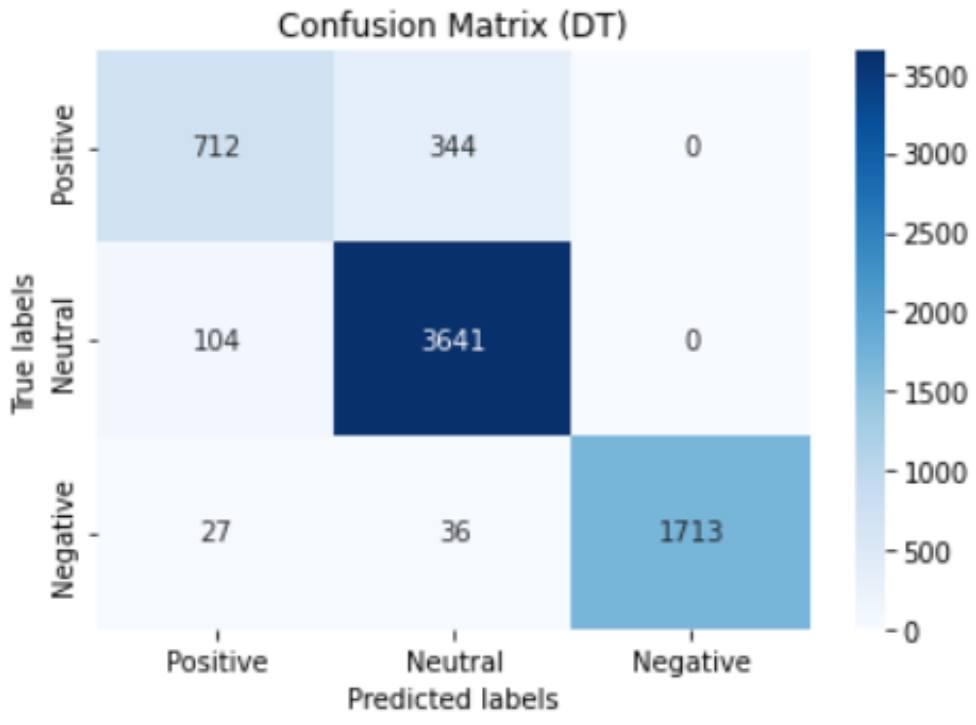

**Fig. 6.** Confusion Matrix of DT

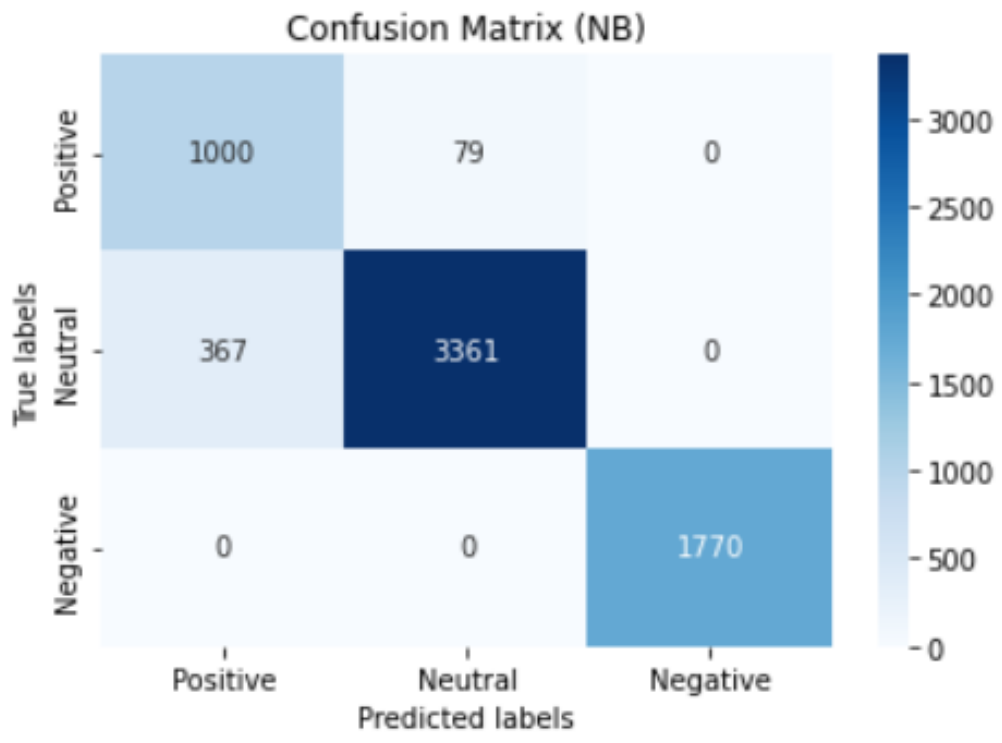

**Fig. 7.** Confusion Matrix of NB

## 6 Conclusion

The research proposes the Twitter Sentiment Analysis of tweets about Israel towards Palestinian-Israeli conflict using advanced technology like Machine Learning. The techniques like Support Vector Classifier (SVC), Decision Tree (DT), and Naïve Bayes (NB) were used to train the model. The result achieved was that Naïve Bayes (NB) obtained the highest accuracy of 93.21%. As part of this research, an attempt has been made to study and implement various algorithms to enhance the feasibility of sentiment analysis where massive amounts of data can be analyzed without much time consumption. The aim is to explore and perform future research on sentiment analysis using Deep Learning Neural Networks. This study, hence, studies all the features and elements, implements them and makes an entirely accurate prediction. Therefore, this will overall contribute to making future decisions about political events.

## 7 Acknowledgement

This research is sponsored by Learn By Research Organization, India.

## References

1. Alamoodi, A. H. et al.; "Sentiment Analysis and Its Applications in Fighting COVID-19 and Infectious Diseases: A Systematic Review."; *Expert Systems with Applications*; 2021.
2. Feldman, Ronen; "Techniques and Applications for Sentiment Analysis: The Main Applications and Challenges of One of the Hottest Research Areas in Computer Science."; *Communications of the ACM*; 2019.
3. Rahmatika et al.; "The Effectiveness of Youtube as an Online Learning Media"; *Journal of Education Technology*; 2021.
4. Tafesse, Wondwesen; "YouTube Marketing: How Marketers' Video Optimization Practices Influence Video Views."; *Internet Research 30.6*; 2020.

5. Bozkurt et al.; "Cleft Lip and Palate YouTube Videos: Content Usefulness and Sentiment Analysis."; *Cleft Palate-Craniofacial Journal 58.3*; 2021.
6. Al-Sarraj et al.; "Bias Detection of Palestinian/Israeli Conflict in Western Media: A Sentiment Analysis Experimental Study."; *International Conference on Promising Electronic Technologies*; 2018.
7. Cambria, Erik; "Affective Computing and Sentiment Analysis."; *IEEE*; 2016.
8. Yadav, Ashima et al.; "Sentiment Analysis Using Deep Learning Architectures: A Review."; *Artificial Intelligence Review 53.6*; 2020.
9. Doaa Mohey et al; "A Survey on Sentiment Analysis Challenges."; *Journal of King Saud University - Engineering Sciences 30.4*; 2018.
10. Rudy, et al.; "Sentiment Analysis: A Combined Approach"; *Journal of Informetrics*; 2009.
11. Chong; "Natural Language Processing for Sentiment Analysis: An Exploratory Analysis on Tweets."; *ICAIET*; 2014.
12. Bhuiyan, Hanif et al.; "Retrieving YouTube Video by Sentiment Analysis on User Comment."; *ICSIPA*; 2017.
13. Novendri, Risky et al.; "Sentiment Analysis of YouTube Movie Trailer Comments Using Naïve Bayes."; *Bulletin of Computer Science and Electrical Engineering 1.1*; 2020.
14. Medhat et al.; "Sentiment Analysis Algorithms and Applications: A Survey."; *Ain Shams Engineering Journal 5.4*; 2014.
15. Dang, Nhan Cach, María N. Moreno-García, and Fernando De la Prieta; "Sentiment Analysis Based on Deep Learning: A Comparative Study."; *Electronics (Switzerland)*; 2020.
16. Phan, Huyen Trang et al.; "Improving the Performance of Sentiment Analysis of Tweets Containing Fuzzy Sentiment Using the Feature Ensemble Model."; *IEEE Access 8*; 2020.
17. Gonçalves, Pollyanna et al.; "Comparing and Combining Sentiment Analysis Methods."; *COSN 2013 - Association for Computing Machinery*; 2013.
18. Do, Hai Ha et al.; "Deep Learning for Aspect-Based Sentiment Analysis: A Comparative Review."; *Expert Systems with Applications*; 2019.
19. Srivastava, Ankit et al.; "Sentiment Analysis of Twitter Data: A Hybrid Approach."; *International Journal of Healthcare Information Systems and Informatics*; 2019.
20. Xu, Guixian et al.; "Sentiment Analysis of Comment Texts Based on BiLSTM."; *IEEE Access 7*; 2019.
21. Al-Agha, Iyad, and Osama Abu-Dahrooj; "Multi-Level Analysis of Political Sentiments Using Twitter Data: A Case Study of the Palestinian-Israeli Conflict."; *Jordanian Journal of Computers and Information Technology*; 2019.
22. Basiri, Mohammad Ehsan et al.; "ABCDM: An Attention-Based Bidirectional CNN-RNN Deep Model for Sentiment Analysis."; *Future Generation Computer Systems*; 2021.
23. Vishal A. Kharde et al.; "Sentiment Analysis of Twitter Data: A Survey of Techniques"; *International Journal of Computer Applications;* 2017.
24. G. Gautam et al.; "Sentiment analysis of twitter data using machine learning approaches and semantic analysis"; *Seventh International Conference on Contemporary Computing (IC3)*; 2014.
25. Abdullah Alsaeedi et al.; "A Study on Sentiment Analysis Techniques of Twitter Data"; *International Journal of Advanced Computer Science and Applications*; 2019.